# A Comparison of Neuroelectrophysiology Databases

## Authors
Priyanka Subash[1], Alex Gray[1], Misque Boswell[1], Samantha L. Cohen[1], Rachael Garner[1], Sana Salehi[1], Calvary Fisher[1], Samuel Hobel[1], Satrajit Ghosh[3], Yaroslav Halchenko[4], Benjamin Dichter[5], Russell A. Poldrack[6], Chris Markiewicz[6], Dora Hermes[7], Arnaud Delorme[8], Scott Makeig[8], Brendan Behan[9], Alana Sparks[10], Stephen R Arnott[11], Zhengjia Wang[12], John Magnotti[12], Michael S. Beauchamp[12], Nader Pouratian[2], Arthur W. Toga[1], Dominique Duncan[1]

## Affiliations
1. Laboratory of Neuro Imaging, USC Stevens Neuroimaging and Informatics Institute, Keck School of Medicine of USC, University of Southern California, 2025 Zonal Avenue, Los Angeles CA 90033
2. Department of Neurological Surgery, University of Texas Southwestern Medical Center, 5303 Harry Hines Blvd, Dallas TX 75390
3. McGovern Institute for Brain Research, MIT Brain and Cognitive Sciences, 77 Massachusetts Avenue, Cambridge MA 02139
4. Department of Psychological & Brain Sciences, Center for Cognitive Neuroscience, Dartmouth Brain Imaging Center, Dartmouth College, 6207 Moore Hall, Hanover NH 03755
5. CatalystNeuro, Benicia CA 94510
6. Department of Psychology, Stanford University, 450 Jane Stanford Way, Stanford CA 94305
7. Mayo Clinic, Department of Physiology & Biomedical Engineering, 200 1st Street SW, Rochester MN 55905
8. Swartz Center of Computational Neuroscience, INC, University of California San Diego, La Jolla CA 92093
9. Ontario Brain Institute, 1 Richmond Street West, Toronto ON M5H 3W4, Canada
10. Indoc Research, Toronto ON, Canada
11. Rotman Research Institute, Toronto ON, Canada
12. Department of Neurosurgery, Perelman School of Medicine, University of Pennsylvania, 3700 Hamilton Walk, Philadelphia PA 19104

## Abstract
As data sharing has become more prevalent, three pillars - archives, standards, and analysis tools - have emerged as critical components in facilitating effective data sharing and collaboration. This paper compares four freely available intracranial neuroelectrophysiology data repositories: Data Archive for the BRAIN Initiative (DABI), Distributed Archives for Neurophysiology Data Integration (DANDI), OpenNeuro, and Brain-CODE. The aim of this review is to describe archives that provide researchers with tools to store, share, and reanalyze both human and non-human neurophysiology data based on criteria that are of interest to the neuroscientific community. The Brain Imaging Data Structure (BIDS) and Neurodata Without Borders (NWB) are utilized by these archives to make data more accessible to researchers by implementing a common standard. As the necessity for integrating large-scale analysis into data repository platforms continues to grow within the neuroscientific community, this article will highlight the various analytical and customizable tools developed within the chosen archives that may advance the field of neuroinformatics.

## Introduction
### Open Science
Open science aims to make research data more transparent and widely available while promoting interdisciplinary partnerships that leverage findings[1,2]. In the United States, public health organizations, such as the National Institutes of Health (NIH), are funding data repositories that can serve as reliable resources to make data accessible. Requiring and encouraging data sharing, promoting common standards, and providing tools for analysis are the foundation of translating research findings into new knowledge, products, and procedures[3–6].

Access to existing datasets presents many advantages, including developing new hypotheses and serving as a source for preliminary analyses. Further, secondary analysis (analysis of existing data to address a different question from the original study) is critical in novel research fields where limited data hinder the production of replicable results, and pooling data sets can add statistical power to an otherwise limited study. Lastly, reused datasets can validate previous conclusions or be repurposed to address new questions with lower cost and effort.

Intracranial electroencephalography (iEEG) provides high temporal and spatial resolution, enabling a level of detail not possible using other neurodata capturing techniques. With iEEG more widely utilized in clinical settings and FDA approvals of deep brain stimulation (DBS) for multiple conditions in the last three decades, electrophysiology studies have become more critical and frequently employed in neuroscience.

**Intracranial Neuroelectrophysiology**
Intracranial neuroelectrophysiology data can be collected through electrodes placed on the cortical surface for electrocorticography (ECoG), intracortical for stereoelectroencephalography (sEEG), or from deep brain stimulation (DBS) electrodes. The recordings are obtained from patients undergoing clinically indicated brain surgery for neurological conditions or those participating in device trials with FDA investigational device exemption (IDE) approval[7–9]. The complexity of the implantation procedures requires multimodal imaging data for proper placement of the electrodes, enriching the resultant datasets. The nature of these procedures, their high costs, and their specialized clinical requirements make these studies relatively rare. In addition, electrodes are placed sparsely, covering different brain regions across patients. Therefore, limited sample sizes[9–11] establish a need for centralized databases to make rare data types available to the larger community for large-scale studies.

**History of Intracranial Neuroelectrophysiology Databases**
So far, there have been several significant developments toward creating valuable databases housing neuroelectrophysiology data. Notable pioneers include EPILEPSIAE (http://www.epilepsiae.eu), iEEG.org (https://www.ieeg.org), EEGLAB (https://eeglab.org ), and Collaborative Research in Computational Neuroscience (CRCNS) (https://crcns.org).

One of the early efforts to construct an electrophysiology repository was undertaken in 2012 by the Epilepsy Research Group at the University of Leuven (Katholieke Universiteit Leuven) in Belgium. EPILEPSIAE (Evolving Platform for Improving the Living Expectations of Patients Suffering from IctAl Events)[12] with a repository subdivision known as The European Epilepsy Database, was developed to provide access to expert-annotated electrophysiology recordings along with metadata and imaging for 275 patients through serving as a paid resource for researchers, clinicians, and students.

Another early platform for data sharing and collaboration, iEEG.org, was established in 2013 by the University of Pennsylvania and the Mayo Clinic. iEEG.org revolutionized the creation and curation of intracranial neurophysiologic and multimodal datasets while making large-scale complex analysis and customization easier for researchers[13].

A different approach was undertaken by the creators of EEGLAB, an environment for human EEG analysis developed at the University of California, San Diego (UCSD)[13] in 2004. EEGLAB gathered contributions from programmers, tool authors, and users while providing access to 32-channel EEG recordings from 14 patients, which later became available on the OpenNeuro platform. In 2019, EEGLAB creators, jointly with OpenNeuro, built the Neuroelectromagnetic Data Archive and Tools Resource (NEMAR)[14].

CRCNS was established in 2002 in collaboration with funding from the National Science Foundation and National Institutes of Health, with the goal of enabling concerted efforts to understand and share neurodata, stimuli, and analysis tools with researchers worldwide. Data available on the CRCNS platform include physiological recordings from sensory and memory systems, as well as eye movement data[15].

### Governing Bodies with Neurodata Sharing Mandates

**The NIH BRAIN Initiative.** In the United States, in 2013, NIH launched the BRAIN (Brain Research Through Advancing Innovative Neurotechnologies) Initiative to advance neuroscience through multimodal, cross-disciplinary, and multi-institutional research, fostering a more integrative approach. Over $1.5 billion has been invested in investigating treatments for brain disorders while advancing research tools and technologies. BRAIN Initiative studies have produced a wealth of neurodata that can further expand our knowledge, making data archives a vital part of its efforts[16].

Presently, several existing BRAIN Initiative-funded neurophysiology repositories collect data to develop new features and expand the size and scope of their systems while sharing broadly with the scientific community. These include Data Archive for the BRAIN Initiative (DABI)[17], Distributed Archives for Neurophysiology Data Integration (DANDI)[18], and OpenNeuro with its partner analysis platform NEMAR[14].

**Ontario Brain Institute.** In Canada, the Province of Ontario recognized the need to improve the diagnosis and treatment of brain disorders, aiming to implement a province-wide integrated approach to research. As a result, in 2010, it established and funded the Ontario Brain Institute (OBI) to create a patient-centered research system, engage the industry, and drive knowledge exchange between researchers, policymakers, and the neuroscience industry[19]. In 2012, OBI launched Brain-CODE, a data-sharing informatics platform, as a crucial part of its efforts to facilitate and maximize the integrative research approach[19].

Each archive aims to improve public health by increasing research transparency through data accessibility, reproducibility, and inter-institutional collaboration. Data from DABI, DANDI, OpenNeuro, NEMAR, and Brain-CODE contributed to numerous publications[20–34] demonstrating their influential impact on neuroscience research.

**National Institute of Mental Health (NIMH) and The National Institute of Mental Health Data Archive (NDA).** Another repository, the NDA (https://nda.nih.gov), is managed by the NIMH for researchers to store, share, and access research data related to mental health. NDA aims to accelerate scientific research and discovery by sharing de-identified and harmonized data across scientific domains (https://nda.nih.gov). It provides a secure platform for researchers to upload and store clinical, neuroimaging, and genomic data, ensuring that datasets are de-identified, sensitive information is encrypted, and strict access controls are in place. Not all data on NDA are publicly available. Some datasets require access requests and approvals by authorized individuals or institutions.

While NDA shares the goal of accelerating discovery through sharing and reanalyzing existing neurodata, it is distinct from DABI, DANDI, and OpenNeuro, which focus on neuroresearch data collected through the BRAIN Initiative-funded projects. Further, NDA focuses on mental health research data and has a cost associated with data deposition, which is intended to cover the maintenance and curation of the archive.

The aim of this review is to describe archives that provide researchers with tools to store, share, and reanalyze both human and non-human neurophysiology data based on criteria that are of interest to the neuroscientific community.

## Methods

Governmental agencies, academic institutions, and patients engaged in research have collectively acknowledged the imperative of sharing scientific data. This imperative is crucial for enhancing transparency and driving research progress, ultimately minimizing the duplication of efforts and resource allocation. Consequently, data archives hold immense potential to revolutionize scientific research by establishing standardized data collection protocols while safeguarding data privacy, security, and long-term preservation.

Data governance is critical in well-established archive management and data asset control. It involves establishing frameworks that dictate how data are collected, stored, accessed, shared, and organized. In the context of data archives, data governance oversees the entire archival process, encompassing data retention policies, security measures, access controls, and data integrity and privacy. Additionally, it facilitates appropriate and controlled data sharing among relevant stakeholders.

To better appreciate the scope of neurophysiology databases and describe optimal user systems, DABI, DANDI, Brain-CODE, and OpenNeuro, jointly with NEMAR (Note: as NEMAR platform is an analysis partner to OpenNeuro archive and does not store independent data, it will be discussed only in the context of neurodata analysis tools), are summarized and compared to assist individuals in the scientific community who have an interest in sharing and accessing human and non-human neuroelectrophysiology data. Inclusion criteria for the selected archives include accessibility to free human and non-human iEEG data variables, integration of open access or controlled access sharing protocols, establishment in North America with global users, and preferred data archives in NIH or OBI-mandated data sharing initiatives.

Though not exhaustive, this review utilizes the following method of assessment to compare databases containing intracranial recordings, focusing on criteria related to data governance frameworks:
- Data Standards & File Formats
- Data Upload Procedures
- Data Download, Access Protocols, and Policies
- Data Storage and Maintenance
- Analytic Tools

To assist with identifying which archive meets the needs of potential data users or data sharers, summary tables of features are provided at the end of each criteria discussion.

**DABI.** Funded in 2018, DABI was created to facilitate and streamline the dissemination of human and non-human neurophysiology data, focusing on intracranial recordings. DABI emphasizes the organization and analysis with investigators who retain control and ownership of their datasets while fulfilling data-sharing directives. Housed at the University of Southern California Stevens Neuroimaging and Informatics Institute, DABI provides innovative infrastructure for interactive data visualization, processing, sharing, and collaboration among researchers[17].

**DANDI.** Funded in 2019, DANDI is a repository that accepts cellular neurophysiology and neuroimaging datasets termed Dandisets[18] for both human and non-human data. The self-service model allows uploading, organizing, and analyzing data with tools provided by the platform, giving users greater control over their data; however, it requires technical expertise to use the platform effectively. Additional features include storage optimizations and tools, allowing investigators to collaborate outside their institutions. DANDI positions itself as a platform for scientists new to secondary analysis. Led by scientists from the Massachusetts Institute of Technology and Dartmouth College, DANDI is designed to aid in the adoption of Neurodata Without Borders (NWB)[35], Brain Imaging Data Structure (BIDS)[8], and Neuroimaging Data Model (NIDM)[36]. Also included are World Wide Web Consortium-Provenance (W3C-PROV) data, metadata, and provenance standards that address data harmonization challenges and promote interoperability[37].

**OpenNeuro.** Funded in 2018 and led by Stanford University, OpenNeuro is one of the largest repositories of human and non-human neuroimaging data[38]. Developed from an earlier version of the platform, OpenfMRI (https://openfmri.org/), OpenNeuro is built around the BIDS specification to simplify file formats and folder structures for broad accessibility, evolving into its current ecosystem of tools and resources. OpenNeuro began supporting iEEG data in 2019 after the modality was incorporated into the BIDS standards (iEEG-BIDS) as an extension[8]. NEMAR is a partner to OpenNeuro for MEG, EEG, and iEEG data (MEEG) and provides additional MEEG tools for datasets made available for public downloads on the OpenNeuro platform, which undergo quality checks and automatic preprocessing. In addition to BIDS, NEMAR uses detailed descriptions of experimental events stored using the Hierarchical Event Descriptor (HED) system[14,39]. Fig. 1 illustrates the iEEG-BIDS folder structure[8].

**Brain-CODE.** Launched in 2012, Brain-CODE[24] is a platform that provides secure informatics-based data sharing, management, and integration with standards that maximize the interoperability of complex neuroscientific human and non-human datasets. In addition to hosting studies that utilize iEEG and Magnetic Resonance Imaging (MRI) data, Brain-CODE collects and shares clinical measures, neuropsychological, omics, sensor, and other data types to facilitate deeper neuroscientific understanding. Brain-CODE includes processing pipelines, notebooks, and virtual desktops to assist with analytics. The platform further promotes academic and industry collaborations for research and discovery.

## Results

**Data Standards and File Formats.** A variety of neurophysiology data modalities (i.e., EEG, MEG, DBS, and iEEG) results in a wide range of formats and structures, leading to challenges in integrating and analyzing pooled data. The lack of standardization of recorded file formats complicates building large-scale datasets and requires file conversion. The emergence of intracranial neurophysiology databases necessitated improved standardization and harmonization protocols to ensure data usability and integration. DABI, DANDI, OpenNeuro, and Brain-CODE offer nuanced solutions to address this demand.

**The Brain Imaging Data Structure (BIDS).** BIDS has gained broad acceptance by the neuroimaging community, becoming the leading standard for harmonizing imaging data. As previously mentioned, electrophysiology data is complex and challenging to harmonize because there are many different formats in which the recording devices store the (source) data. Several electrophysiology data formats are allowed in the BIDS specification. For EEG, these include European Data Format and its extensions (EDF/EDF+/BDF)[40], Brain Vision Core Data Format[41], and EEGLAB. iEEG additionally allows constrained NWB and MEF3 files to allow data chunking (NWB & MEF3), lossless compression (NWB & MEF3) and HIPAA-compliant multi-layer encryption of sensitive data (MEF3). Lastly, MEG is limited to CTF, Neuromag, BTi / 4D Neuroimaging, KIT/Yokogawa, KRISS, and Chieti file formats (Note that this is not a strict rule, as some iEEG and MEG files may contain EEG channels). While there are some differences in the formats across modalities, the overall structure is harmonized such that metadata with information about channels, electrodes, and events are stored similarly across MEG, EEG, and iEEG modalities.

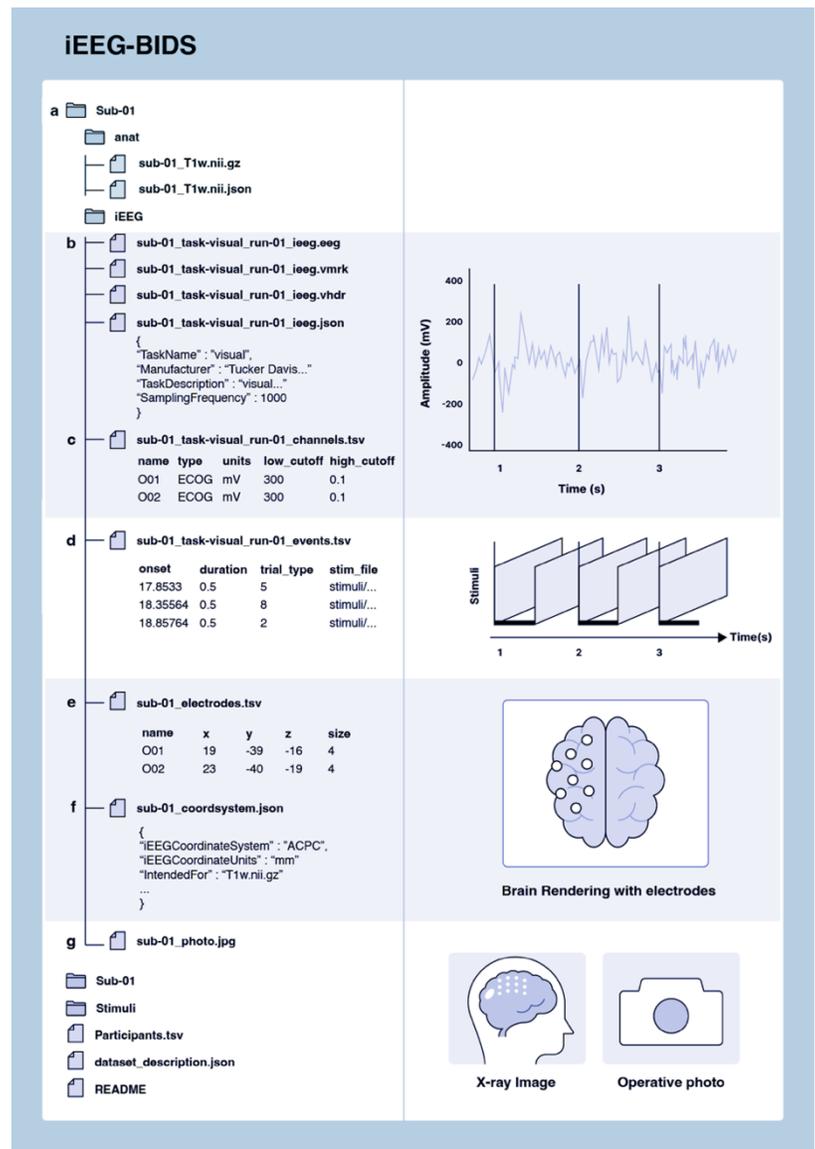

**Fig. 1** iEEG-BIDS folder structure.
**(a)** BIDS structure contains folders for each subject and one folder for stimuli. Within a subject folder, an /anat/ folder contains structural images alongside iEEG data. **(b)** _ieeg.json file stores iEEG data containing information on acquisition systems and their parameters. **(c)** _channels.tsv file stores metadata about channel-specific information, such as hardware filters or electrophysiological units. **(d)** _events.tsv TSV file contains event timing data. **(e)** _electrodes.tsv files store electrode coordinates. **(f)** _coordsystem.json file stores the coordinate system information. **(g)** Other images relevant for iEEG, such as surface models and 2-D images can be stored in a systematic manner. Optional folders and labels, such as the session folder and space- label, are mostly left out of this example.

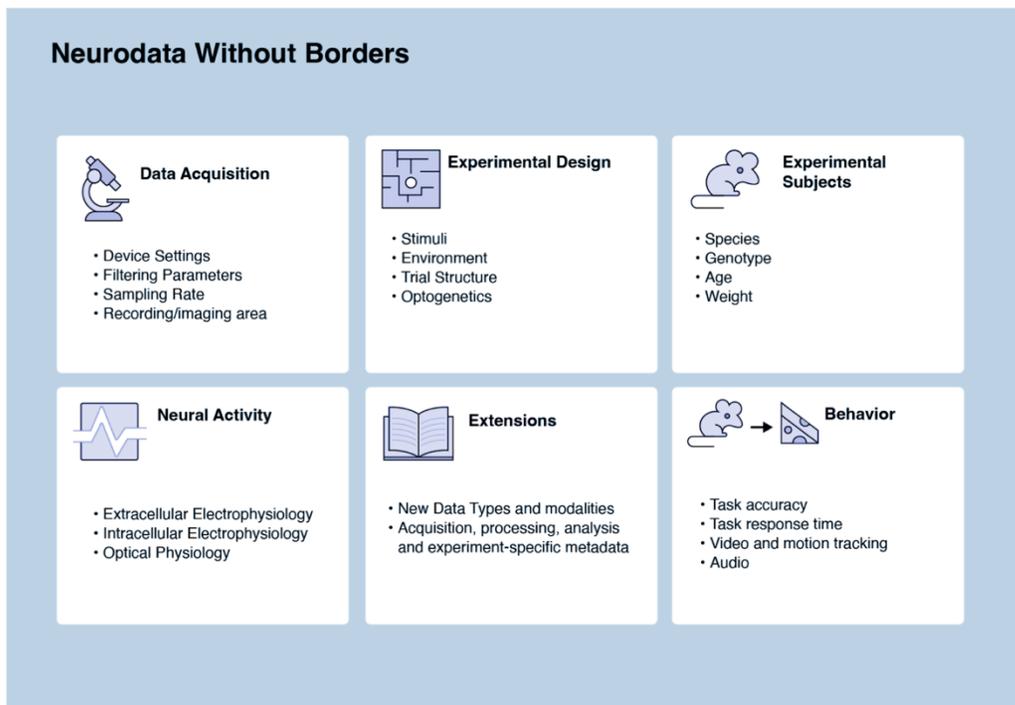

Fig. 2 NWB Data Types.

**Neurodata Without Borders (NWB).** NWB format is a standard that packages neurophysiology data with the metadata necessary for reanalysis. NWB is primarily used for cellular neurophysiology data such as extra- and intra-cellular electrophysiology, optical physiology, and behavior (Fig. 2). Several NWB datasets on DANDI and DABI contain iEEG data[34], but it is not commonly used for EEG or MEG. In contrast to BIDS, which supports storing acquired data in domain-specific formats, NWB requires that the electrophysiology measurements be stored within the NWB file. Although this creates a higher barrier for data conversion, it provides increased standardization and enables advanced data engineering tools such as data chunking and lossless compression.

Each repository discussed here approaches data standards differently. Some archives place the burden of file conversions on the data providers. Others take on the task themselves or leave the harmonization protocol to the data users to decide and execute.

**DABI.** DABI hosts a broad range of multimodal data emphasizing intracranial neurophysiology. Neurological diagnostic test and procedure subtypes, imaging, behavioral, demographic, and clinical variables are also stored on the platform. Modalities of data include iEEG, EEG, electromyography (EMG), single/multi-unit microelectrode recordings, DBS, MRI, fMRI, DWI, positron emission tomography (PET), and computed tomography (CT). DABI accepts multiple data formats (see Table 1 for a comprehensive list) to alleviate the challenge of time-consuming file conversions but strongly encourages using NWB and BIDS standards when possible. The variety of data formats and modalities within DABI is intended to be all-encompassing and includes scripts from Python[42], MATLAB[43], and R[44]. Users are free to upload either raw or processed forms.

**DANDI.** DANDI houses neurophysiology data, including electrophysiology, optophysiology, microscopy (e.g., selective plane illumination microscopy (SPIM)[45]), behavioral time series, and imaging data (e.g., MRI, fMRI, OCT). The platform requires that uploaded data adhere to established community standards and file formats. The individual file level includes NWB for cellular neurophysiology and optical physiology and OME-Zarr implementation[46] of the Next-generation file format (NGFF)[47] for microscopy (see Code Availability: 1). The entire dataset (Dandiset in DANDI terms) must adhere to BIDS-like lightweight file tree hierarchy as prepared by dandi command line tools. For cellular neurophysiology, such as electrophysiology and optical physiology, DANDI requires the NWB format. BIDS is required for neuroimaging data, encompassing structural MRI, fMRI, and microscopy data, while the NIDM standard is used for associated metadata[35,36,45–47]. The DANDI team is actively participating in BIDS and NWB standards developments to ensure that those adequately address the needs of the archive users' use cases.

**OpenNeuro.** OpenNeuro focuses on collecting imaging and electrophysiologic data modalities. Imaging data include structural MRI, fMRI, and PET. Electrophysiologic data include EEG, iEEG, and MEG. All data uploaded onto OpenNeuro are submitted through a BIDS-validator and must be BIDS-compliant to be included. NEMAR is the first open data archive to implement the Hierarchical Event Descriptor (HED) tags for data discovery and integration[14]. The HED system provides a standardized and flexible set of descriptors for experimental events in brain imaging or behavioral experiments and has been integrated into BIDS as a standard for describing events[14,39].

**Brain-CODE.** The primary focus of Brain-CODE is to collect multidimensional neuroscience data involving all forms of Magnetic Resonance Imaging (MRI), ocular computed tomography (OCT), genomic next-generation sequencing (NGS), proteomic, and electrophysiological data, along with clinical, wearable devices, and other data types. Accepted file formats include tabular (e.g., csv, txt), imaging (e.g., NifTI, DICOM), molecular (e.g.,vcf, fastq), and MATLAB files (.mat).

Brain-CODE utilizes modality-specific electronic data capture systems for collection and management. These include REDCap (https://www.project-redcap.org/) or OpenClinica[48] for clinical data, Stroke Patient Recovery Research Database (SPReD) – a system built on the eXtensible Neuroimaging Archive Toolkit (XNAT)[49] – for imaging data, LabKey (https://www.labkey.com) for molecular data, and a custom-built Subject Registry for the secure collection of encrypted provincial health card numbers. Further, Brain-CODE provides data quality and processing pipelines, a central data federation system enabling cross-modality, site, and study data federation, data processing and analytic-driven workspaces, data query, and visual data analytics solutions.

| Feature | DABI | DANDI | OpenNeuro | Brain-CODE |
|---|---|---|---|---|
| BIDS | Yes | Yes (imaging only) | Yes | Yes |
| NWB | Yes | Yes | | |
| NIDM | | Yes | | |
| DICOM | Yes | | | |
| NifTI | Yes | Yes (as part of BIDS) | Yes (as part of BIDS) | Yes (as part of BIDS) |
| MATLAB | Yes | | | Yes |
| BrainVision | Yes | | Yes (as part of BIDS) | Yes (as part of BIDS) |
| EEGLAB | Yes | | Yes (as part of BIDS) | Yes (as part of BIDS) |
| BioSemi | Yes | | Yes (as part of BIDS) | Yes (as part of BIDS) |
| European Data Format (EDF) | Yes | Yes (as part of NWB) | Yes (as part of BIDS) | Yes (as part of BIDS) |
| Blackrock NeuroPort | Yes | Yes (as part of NWB) | | |
| Intan | Yes | Yes (as part of NWB) | | |
| JSON (for data) | Yes | | | |

**Table 1** Data Standards and File Formats.

**Data Upload Procedures.** DABI, DANDI, OpenNeuro, and Brain-CODE employ distinctive methods for data upload with some overlap between platforms. The challenges of high-speed data migration are met with software proxies that assist in streamlining the process for smoother transitions and provide versatility to accommodate the needs of the providers.

**DABI.** DABI offers users four options for uploading (Fig. 3). The first is through a browser-based "DABI Web Uploader" within the DABI portal, where providers can directly upload data to their affiliated DABI projects. The Web Uploader option is preferred for small files (up to 50GB) due to dependency on a continuous internet connection. If a session times out during a single file upload, the process will need to be restarted. However, should an error occur with multiple file uploads, those successfully uploaded up to that point will remain on the server.

The second upload option is through IBM's file transfer client Aspera (https://www.ibm.com/products/aspera), installed locally. Aspera is an encrypted, HIPAA-compliant, high-speed file transfer system that is 10 to 100 times faster than the standard file transfer protocol and is primarily used for large, raw, or preprocessed datasets. Aspera, which requires a separate account, can connect to DABI servers for uploading or downloading data, granting its users complete control over their data and providing a safety net for timeouts during uploads by resuming from its last point before the cutoff.

The third alternative is the SSH File Transfer Protocol (SFTP), an upload method for large raw or preprocessed datasets. DABI chooses to support this method due to potential approval restrictions for new software and presently established SFTP protocols by several institutions.

The final option is through a HIPAA-compliant cloud-connected Box account (https://www.box.com/home), allowing real-time data collection. Updates will automatically sync if a provider has already uploaded data to Box, eliminating the need to upload through another method. Files stored in a cloud account are not transferred to DABI servers by default. Instead, DABI facilitates transferring files between the cloud and the user's machine.

**DANDI.** Providers utilizing DANDI can register on the portal with a GitHub account, create a new dataset, and, using the Python-based DANDI CLI tool, organize their collection of NWB files into a conformant file tree, validate adherence of metadata to standards and DANDI requirements, and then upload (Fig. 4). The upload can be performed using the same dandi CLI or the corresponding DANDI Python library or directly communicating with a DANDI archive server through a REST application programming interface (API). A staging instance of the DANDI archive can be used before uploading to the main archive for experimentation or testing of the automations. The DANDI Python library is also often used to automate tasks, perform data analysis, or interact with other software.

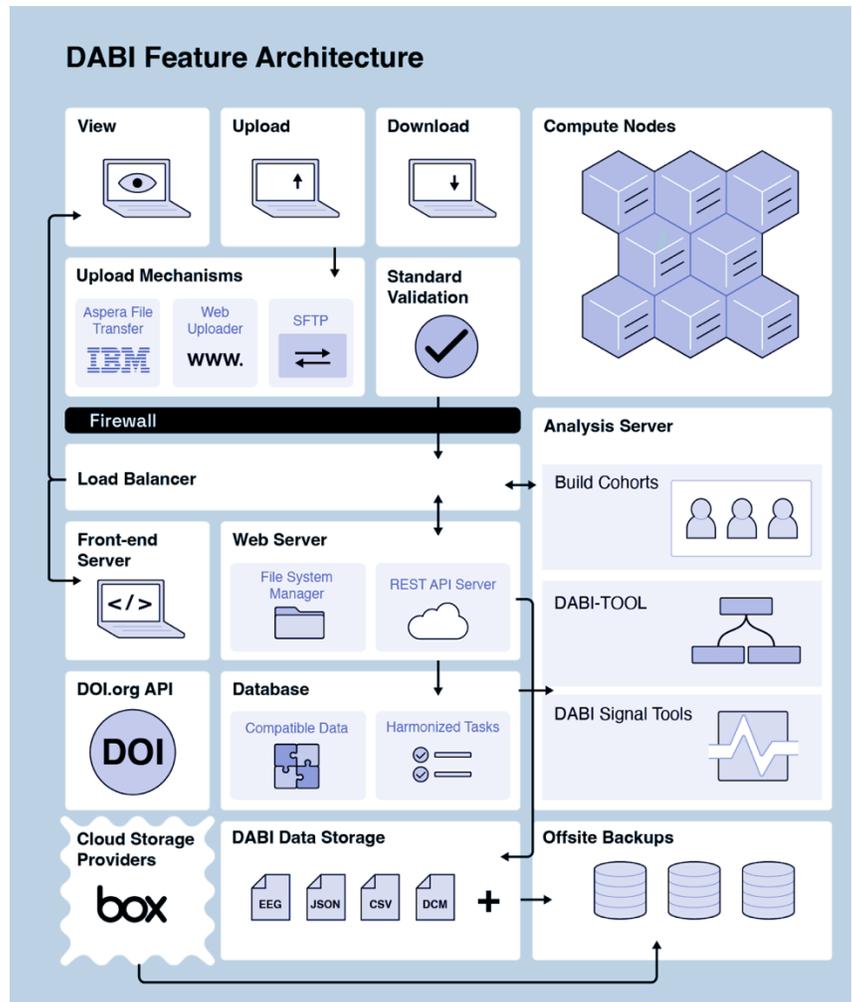

**Fig.3** DABI Architecture

Once a data collection - or Dandiset - is registered, a unique identifier will be generated that can be referenced to initiate any Dandiset-level metadata changes on the platform. Because DANDI requires adherence to standards (e.g., NWB and BIDS), uploaded data are validated for format errors. If errors are found, users must convert their files into the standard formats or correct coding issues when running within the CLI before successful data deposition. DANDI provides a comprehensive online handbook, DANDI CLI documentation, and YouTube tutorials that outline each step of the upload process. Datasets can be "embargoed" to be made accessible only via authorization by the provider. They remain unavailable to public view, allowing researchers to solicit feedback before publishing findings or sharing them without concern about intellectual property disclosure. A dataset passing validation can be "published" as a versioned release on the archive, which is then assigned a datacite DOI, and guarantees future availability of this particular version of the dataset.

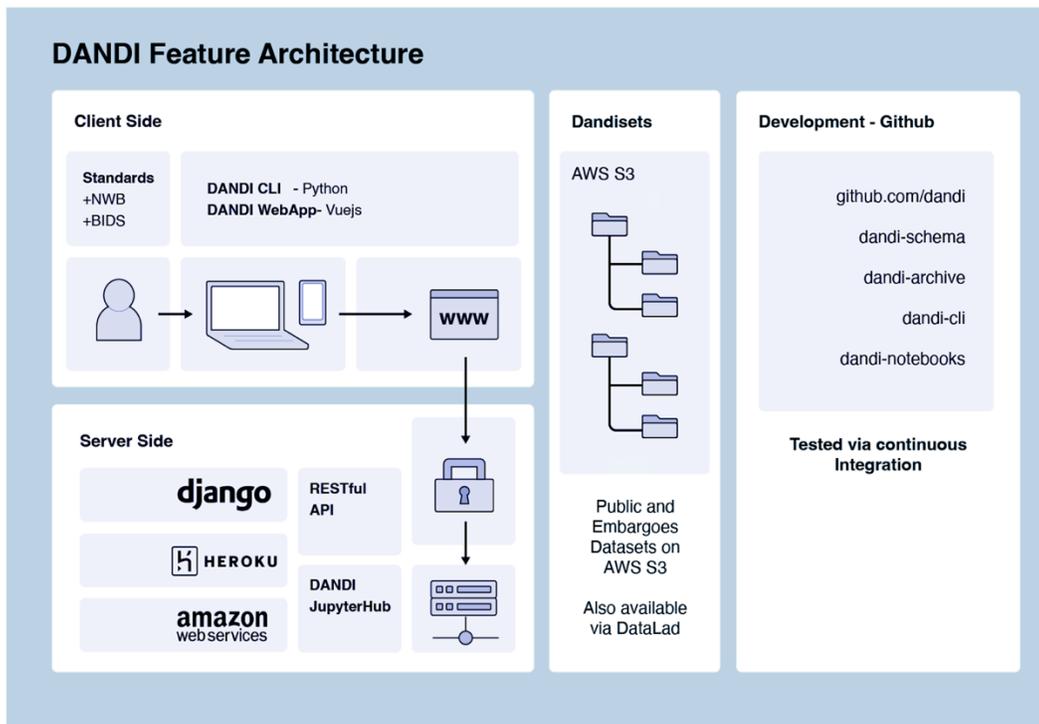

**Fig. 4** DANDI Feature Architecture.

**OpenNeuro.** OpenNeuro provides collaborators with two options for uploads. The first is similar to DANDI's CLI tool, allowing partners to upload data through their in-house CLI tool. Since uploads are read through a BIDS-validator before data deposition, collaborators who use this method must ensure that the files are BIDS-compliant. The second upload option is a web interface tool similar to DABI's Web Uploader. Within the OpenNeuro user interface, users select files that undergo a validation process to ensure BIDS compliance before upload (Fig. 1)[38]. Additional constraints are placed on datasets, such as a non-empty author list when required to ensure that datasets are findable and citable.

Once uploaded, datasets are assigned a unique accession number and enter the "draft" state. Uploaders and authorized users may upload additional files through the CLI tool or the dataset's landing page. When a dataset passes validation, a "snapshot" may be made, which assigns a version number and a digital object identifier (DOI) to that snapshot. The dataset is then in an embargoed state. When the dataset is published, either by the uploader or automatically after the 36-month embargo period, all versions of the dataset are made public (Fig. 5).

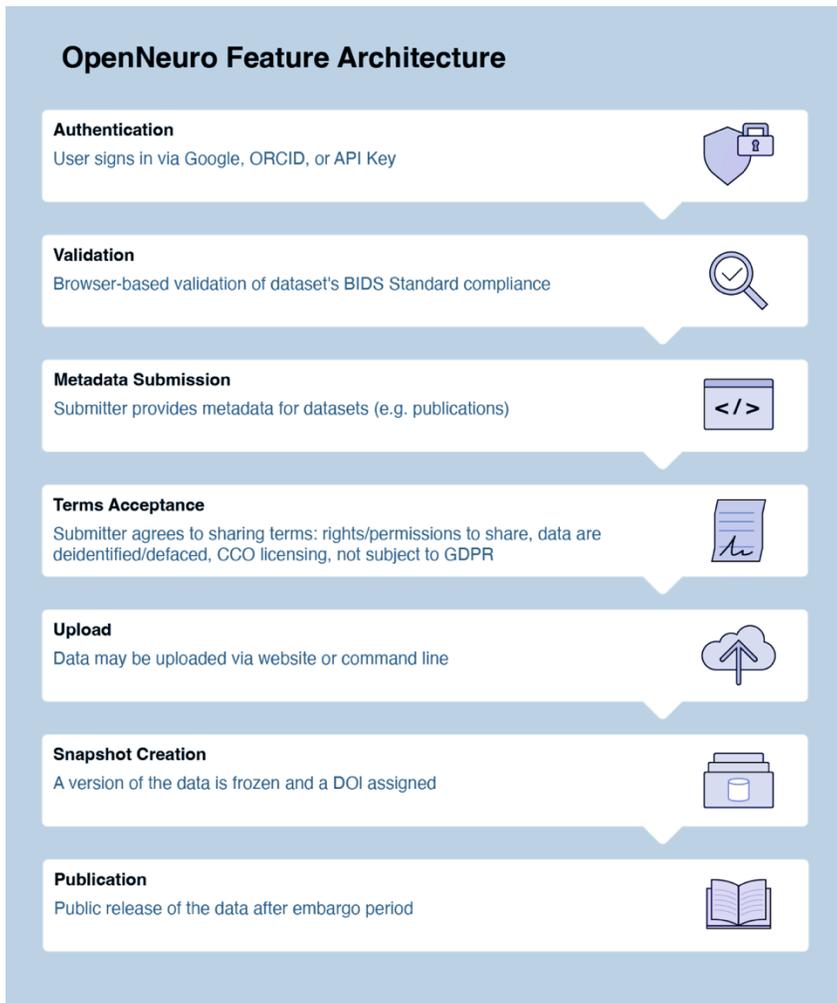

**Fig. 5** OpenNeuro Data Uploading Process Flow.

**Brain-CODE.**
Brain-CODE supports studies requiring data collection from several sites across multiple modalities; thus, collection workflows must provide sufficient flexibility (Fig. 6). Each modality can be collected independently without the platform enforcing a specific workflow. Imaging and other data can be submitted via a secure web browser by manual or bulk transfer into XNAT. Clinical, genomic, and other "omic" data can be shared through electronic data capture systems such as REDCap and LabKey. Data uploaded via data capture tools are ingested, processed, and merged into the central Brain-CODE Federation System daily. The data collected across multiple modalities are organized using a standard participant-naming convention and undergo administrative and quality control processes, study tracking and monitoring dashboards, data query, and release and data analytics.

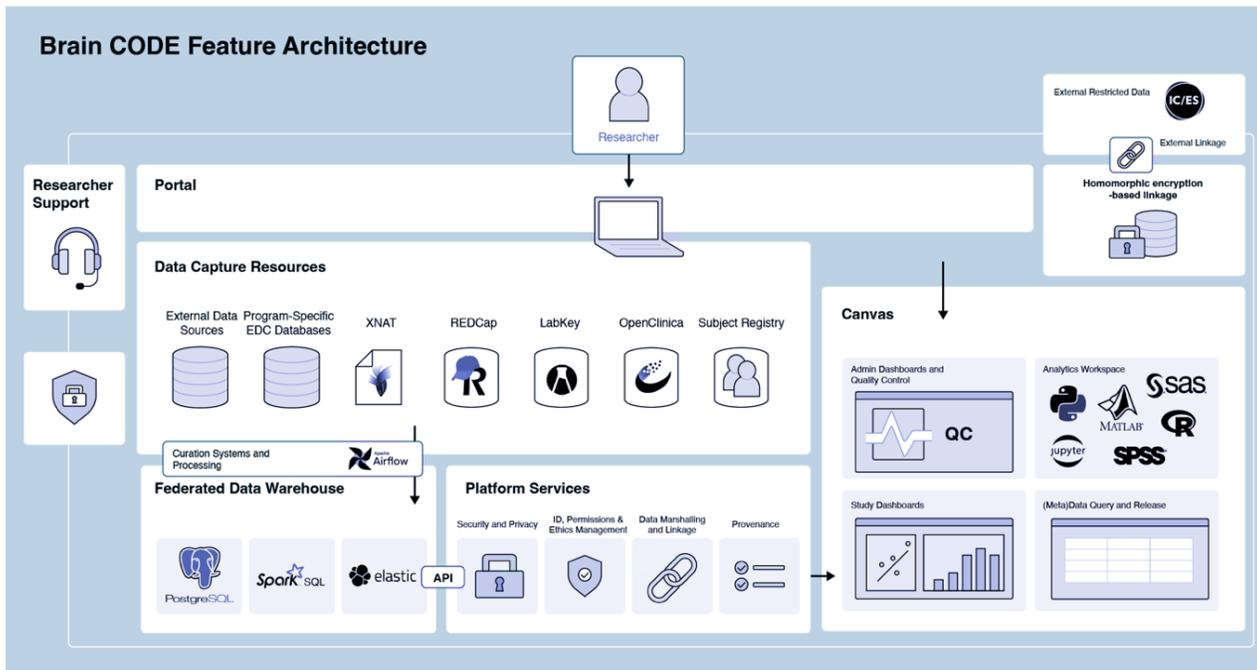

**Fig. 6** Brain-CODE Feature Architecture.

| Feature | DABI | DANDI | OpenNeuro | Brain-CODE |
|---|---|---|---|---|
| Multiple options for upload | Yes | | Yes | |
| Cloud linking (no upload) | Yes | | | |
| CLI Upload | | Yes | Yes | |
| Web Upload | Yes | | Yes | Yes |

**Table 2** Data Upload Procedures.

**Data Download And Access Protocols.** Download and access protocols for intracranial neurophysiology data are critical to promoting data sharing and research collaboration. Protecting data and authentication information as it travels through networks or between servers is paramount to preserving data integrity on each platform. DABI, DANDI, OpenNeuro, and Brain-CODE data are subject to specific protocols that reflect each platform's high usability standards.

**DABI.** DABI download procedures require all users to create an account and sign a DABI Data Use Agreement (DUA). This allows them to download the data directly from DABI if a data owner's project is listed as public. For private datasets, access must be granted by the data owners, sometimes requiring additional institutional DUAs with terms particular to each dataset. DABI's provider portal facilitates communication between users requesting data access and data providers. For a more granular approach to securing data, owners can control access to individual subject data or files by creating subprojects with specific sharing parameters.

**DANDI.** DANDI allows users to download public datasets via five methods without creating an account (access to embargoed data requires an account and permission for a given dataset).

The first is through the DANDI web application. Once data users have identified a Dandiset of interest, they can navigate to the associated Files icon, visualize the data like a directory tree, and download any files. This method is convenient for inexperienced users needing to download only a few files.

The second, more flexible method utilizes the Python library and provides a CLI. Data users must first download the Python client and then download the Dandiset (single files, a set of files, a single subject, or the entire Dandiset) using DANDI CLI or Python API within the Python environment. The CLI tool is the default recommended method for a typical user to download data.

The third method uses DataLad[50], an open-source git[51] and git-annex (https://git-annex.branchable.com/) based tool for flexible data management with CLI and Python interfaces. Like OpenNeuro, DANDI exposes all Dandisets as versioned DataLad datasets from GitHub (see Code Availability: 4), allowing users to view an entire DataLad set without downloading any data in their local file system and permitting them to download specific files and folders that fulfill their needs.

Moreover, using the modularity principle of git submodules, such DANDI DataLad datasets could be combined within larger study datasets and accompanied by computational environment containers to facilitate reproducible computation and collaboration. This method is recommended for users aspiring to keep all of their data under version control and often provides a more convenient approach for navigation and selective downloads within datasets.

The fourth method uses an API server directly where every file can be downloaded via simple REST API. This method is recommended for non-Python (e.g., MATLAB) developers to provide integrated solutions with the DANDI archive.

And finally, it is possible to access data directly on the underlying S3 bucket using AWS[52] tools or the HTTP protocol. For every Dandiset version, DANDI provides dumps of listings of assets. Direct access via HTTP to the S3 bucket allows on-demand data access using Range requests, e.g., as implemented in fsspec and ROS3 driver for HDF5. This method is recommended for efficient remote access to metadata and data within NWB files, OME-Zarr file sets, and archival.

**OpenNeuro.** Four download mechanisms are available to OpenNeuro users: the web interface, the OpenNeuro CLI, DataLad, and the Amazon S3 client. The web interface is available to any user with permission to view the dataset, including users without accounts in the case of public datasets; through this interface, the whole dataset or individual files may be downloaded from any accessible version. Dataset owners may download data from unpublished drafts. The OpenNeuro CLI requires logging in with an API token. Versioned or draft datasets that the user has access to may be downloaded through this mechanism. This mechanism supports resuming interrupted downloads in case of large datasets and unstable connections. OpenNeuro publishes mirrors of its public datasets to the OpenNeuro Datasets organization on GitHub, providing public DataLad access to the full version history up to the latest snapshot. In addition, dataset owners have read/write DataLad access to datasets through private URLs that require API tokens, which further provide access to draft or embargoed data (see Code Availability: 2). The Amazon AWS client may be used to download the latest version of a public dataset. Once an EEG, MEG, or iEEG dataset is made public on OpenNeuro, within 24 hours, its clone automatically appears on NEMAR. Unlike OpenNeuro, its partner archive, NEMAR, offers only full-archive ZIP file downloads, allowing users to bypass OpenNeuro's Amazon cloud server and achieve significantly faster speeds through transfers from the San Diego Supercomputer Center[14].

Like DANDI, OpenNeuro utilizes the open-source command line tool DataLad to manage, organize, and share research data to facilitate reproducibility and collaboration. DataLad uses a distributed version control system (Git) as the underlying technology, allowing researchers to track their data over time and collaborate with others by sharing and merging changes. DataLad also provides tools for managing metadata and software. One of the most appealing features is its ability to manage data dependencies. Researchers can specify the data required for analysis and automatically retrieve the sets from remote locations, including data repositories or datasets created by other researchers with DataLad.

**Brain-CODE.** Brain-CODE requires all users to create an account prior to requesting datasets through their platform, which provides access to data via a tiered 3-zone process.

Zone 1 holds raw data supplied to OBI by affiliated providers. Data funneled into Zone 1 may contain personal identifiers and is considered "raw" until de-identified. It is primarily accessible to the providers, who must ensure that it aligns with the Research Ethics Board (REB) protocols. De-identified data can be shared with external collaborators through a Data Use Agreement (DUA) with appropriate REB approvals. Zone 2 focuses on long-term data storage of de-identified data with public and controlled access options for third parties. Since the datasets have already been integrated in Zone 1 with the Brain-CODE federation system, metadata is available to external researchers and can be queried via Data Request Portal dashboards. Data are disclosed to external researchers in Zone 3 in alignment with their data access request and data use agreement.

There are two access mechanisms in Zone 3: Public and Controlled Access. Users can immediately download the data if they submit a Public data access request. Controlled access requests are more nuanced and require OBI and its Data Access Committee's review to ensure compliance with the Brain-CODE Informatics Governance Policy, applicable laws and guidelines. Once the user's access is approved, the data can then be analyzed within Brain-CODE Workspace or downloaded locally, following the completion of a Data Transfer Agreement[53].

Secure data protocols afforded by each archive's role-based (RBAC) or attribute-based (ABAC) access controls give data providers authority over users' access and download permissions. OpenNeuro and DANDI permit data variables to be publicly available immediately or after a 36-month grace period. DABI enables data owners to publicize full or partial datasets at their discretion, offering a granular data-control approach. Brain-CODE institutes a three-tiered zoning framework for data access and, like DABI, creates strict access protocols that enable greater control by data owners.

| Feature | DABI | DANDI | OpenNeuro | Brain-CODE |
|---|---|---|---|---|
| Data Use Agreement (DUA) | Yes | Yes (implicit) | Yes | Yes |
| User Account Required | Yes | Yes | | Yes |
| Public Data Sets Available | Yes | Yes | Yes | Yes |
| Private Data Sets Available | Yes | Yes | Yes | Yes |
| Embargo Period | | Yes | Yes | Yes |
| CLI Download | | Yes | Yes | |
| Ability to download part of dataset | Yes | Yes | Yes | Yes |

Table 3 Data Download and Access Protocols.

**Data Storage And Maintenance.** As datasets continue to grow, the roles of storage and management become increasingly important for security, operation, and compliance needs, which necessitate multiprotocol strategies. Information technology (IT) administrators at DABI, DANDI, OpenNeuro, and Brain-CODE apply methods that aim to improve performance and recovery, protect against data loss, avert human error, and thwart data breaches or system failures.

To ensure patient protection, DANDI, DABI, and OpenNeuro follow the National Institutes of Health (NIH) BRAIN Initiative and the NIH Office of Science Policy's (OSP) guidelines for patient privacy.

Privacy and confidentiality of research participants are protected by requiring that any data collected during a study is subject to appropriate security measures, such as encryption and access controls. The data must be de-identified to remove any information that could be used to link them back to a specific individual. Each archive provides stringent measures to ensure policy compliance and protect patients from privacy violations (Note: These policies are more applicable for U.S.-based platforms vs. Canadian Brain-CODE).

**DABI.** DABI permits data providers to choose from three storage options: centralized, federated, and cloud-based. Under the centralized model, data are housed at the University of Southern California Neuroimaging and Informatics Institute – this is the most prevalent method. The federated model permits local storage at the owner's affiliated institution. Under a cloud-based model, users with access permissions can securely communicate with the cloud provider via DABI's central server-cached memory. Under this model, no data are ever written to the DABI disks. Box is the only presently supported cloud service offered by DABI. DABI data maintenance includes manual and automated quality checks as well as performance monitoring to identify bottlenecks, slow-loading pages, or other performance-related issues.

**DANDI.** DANDI harnesses Amazon Web Services (AWS) S3 cloud-based storage services to store public or embargoed Dandisets. Data are maintained through the DANDI web interface and accessible via DataLad. Owners can edit, view, share, and publish their Dandisets directly through the DANDI website, maintaining full ownership of their data throughout the collaboration (Fig. 5). The publishing process generates a DOI for each dataset. Owners can extend the Dandiset with new files by releasing its multiple versions.

**OpenNeuro.** OpenNeuro stores and manages all data through the Google Cloud Platform, with access via the OpenNeuro web interface, CLI tool, and DataLad. Partners maintain data through the OpenNeuro web server to ensure long-term data security[38]. On publication, dataset contents are pushed to Amazon S3, and the meta-data is mirrored to GitHub for full public access.

**Brain-CODE.** Data hosted by Brain-CODE are stored on high-performance computing nodes at the Centre for Advanced Computing (CAC) at Queen's University in Kingston, Ontario, Canada. Data are federated via standard participant IDs providing a better overall data management structure and can be mapped to encrypted provincial health card numbers. The Brain-CODE Subject Registry sub-system encrypts the health card number within the data provider's web browser. The original value (i.e., health card number) of the health card never leaves the research site.

Instead, sensitive information is encrypted using a homomorphic algorithm and transmitted and stored in the Subject Registry. Subsequently, encrypted health card numbers link study data to external databases, such as one holding provincial health administrative data. For data maintenance, study teams are provided with management tools that can be accessed through the web interface to upload new data to an existing project and make edits or changes to existing datasets.

| Feature | DABI | DANDI | OpenNeuro | Brain-CODE |
|---|---|---|---|---|
| Versioned Datasets |  | Yes | Yes | Yes |
| Federated Storage | Yes |  |  |  |
| Centralized Storage | Yes |  |  | Yes |

**Table 4** Data Storage and Maintenance.

**Analytic Tools.** Analytic tools for neurophysiology archives are valuable and convenient resources that enable server-based data analysis, preserve computational resources, and save significant time. While some platforms employ more intricate on-site analysis tools, others may prefer to limit or outsource these functions. With researchers having outside access to a broad collection of analysis software, providing the options already available externally may be expensive and complex, especially considering the broad range of research and the number of data modalities. Some tools and environments, such as Jupyter that have been adopted by the neuroimaging community are more cumbersome to employ with electrophysiology data as they are not designed for processing and analyzing high-frequency electrophysiology signals and are not well optimized for processing large amounts of data from multiple electrodes simultaneously (see Code Availability: 5). Still, accessing tools and running analysis software on the archives' servers without downloading large datasets is an appealing option to many researchers who value time and efficiency.

**DABI.** DABI is working on enhancing in-house analytical tools that allow in-depth analysis of the repository's neurophysiology data. The integration of data analysis tools on DABI is still evolving, with plans to offer a comprehensive, user-friendly pipeline that provides users with server space for complex analyses. The cohort feature on DABI will allow data exploration and aggregation for seamless cross-study and cross-modality cohort building. The pipeline tool will include code and no-code workflow interfaces for processing, analysis, and visualization of MRI and electrophysiology data to accommodate a spectrum of users from novice to advanced. Further, pipeline workflows and preprocessing parameters are designed for sharing with others and saving for subsequent analyses.

DABI features will also include The Jupyter Notebook Library, enabling users to create custom notebooks for comprehensive data analysis. Jupyter and DABI plan to construct a public library where users can communally pool their notebooks and search for notebooks created by others.

DABI partners with RAVE: Reproducible Analysis and Visualization of Intracranial EEG Data[54], a comprehensive package that provides an easy-to-use software platform for all steps of iEEG analysis, from electrode localization to sophisticated statistical analysis of group data. RAVE was designed with a client-server architecture so resource-heavy tasks (such as storing and analyzing terabytes of iEEG data) can occur in the cloud. In contrast, visualization of analysis results (such as interactive manipulation and inspection of iEEG data using a 3-D cortical surface model) occurs locally for a satisfying user experience. Upcoming plans for DABI include pipelines for automated analysis once the appropriate parameters have been determined. Future development of RAVE is focused on easy upload, download, and analysis of data in archives, including DABI, DANDI, and OpenNeuro.

**DANDI.** DANDI provides a Jupyter environment named DandiHub for users to interact with the data. While using DandiHub, a researcher can perform many different kinds of analyses and visualize data; it is not intended for significant computation. Presently, each user is restricted to 48 cores and 96GB of RAM or a machine with a T4 GPU. For additional resources, a researcher can use their own AWS account to extend the compute resources. DANDI web UI integrates with several external services (e.g., MetaCell NWB explorer[35], BioImagSuite/Viewer[55], itk/vtk viewer (see Code Availability: 6-8) for viewing individual files and/or performing analytics.

**OpenNeuro.** Conversely, OpenNeuro does not house analysis tools; however, it partners with external providers for analytics, which act as intermediaries. To perform large-scale analyses, users can process public data through independent platforms, including BrainLife.io (https://brainlife.io/about/) and NEMAR.org. NEMAR allows searching OpenNeuro's dataset metadata to aid data selection; researchers can view a range of data statistics, measures, and transforms. Once the data are selected, users can download pointers to the selected datasets that can be included in scripts uploaded to the Neuroscience Gateway for no-cost processing. Users in various neuroscience computing environments can build NSG data analysis scripts. Importantly, NEMAR academic users, in particular, can use a tight integration via its Neuroscience Gateway (NSG) portal[56] plug-in by starting Python or MATLAB scripts on the Expanse supercomputer that directly interface NEMAR-stored BIDS datasets on the same back end[14].

**Brain-CODE.** Brain-CODE offers access to secure customizable computing workspaces for analysis with tools such as RStudio (https://www.rstudio.com/) and Jupyter Notebooks[57], among others. Researchers also can implement their software licenses in Brain-CODE workspaces; as a result, key licensed tools such as MATLAB, SAS[58], and SPSS[59] can be installed in a remote desktop workspace environment on either Linux or Windows operating systems. Data can be extracted from source data capture tools or the federation system and provided to research teams in approved project-specific workspaces. Researchers can request computing resources as their needs change.

| Feature | DABI | DANDI | OpenNeuro | Brain-CODE |
|---|---|---|---|---|
| Signal Processing Analyses | Yes | | Yes (NEMAR) | |
| Machine Learning Analyses | Yes | | | |
| RAVE | Yes | Yes (DANDIhub) | | |
| Python Notebooks | Yes | Yes (DANDIhub) | | Yes |
| Multi-modal analytics | Yes | | Yes (NEMAR) | Yes |
| On-Cloud Analysis | Yes | Yes (DANDIhub) | Yes (NEMAR) | |
| Signal Processing Analyses | Yes | | | |

**Table 5** Analytic Tools.

## Discussion

**Conclusion**: Public health organizations strive to improve neuroscience research by encouraging data sharing to promote secondary analyses and the translation of research outputs into new information, products, and procedures. As a result, NIH has created specific Requests for Applications (RFAs) to support and promote secondary analysis of existing datasets. United States' DABI, DANDI, OpenNeuro, NEMAR, and Canadian Brain-CODE are government-funded projects based in North America that facilitate data sharing within the scientific community. Each of the archives implements different data governance strategies and frameworks to build a platform that addresses the needs of its providers and users.

DABI took on the challenge of providing the broadest range of analysis tools, while OpenNeuro has elected to employ an independent analysis platform on its partner archive, NEMAR. Both of the solutions carry merits and appeal to researchers. Those who prefer to complete their analyses on the cloud and do not feel limited by the selection of tools will find DABI useful, and those who prefer to use highly specialized tools or machine-learning environments may select other platforms to conduct the analyses. The cost of maintaining and developing data analysis tools and software licensing permissions is an important consideration for the repositories, who need to weigh the frequency of use against the financial investment. Ultimately, while significant resources are necessary to address the challenges of advanced data analytics, the investment is well worth the effort, given the productivity gains and possible experimental breakthroughs downstream.

The first step of the preprocessing pipeline involves harmonization protocols to ensure dataset usability. However, harmonizing neuroelectrophysiology data remains a significant challenge, potentially hindering secondary analysis. Unstructured (unstandardized) files have limitations in pooling, preprocessing, and analysis, which restricts the research potential of these datasets.

Some archives strictly adhere to standardization protocols, while others offer more flexibility in accepted data formats. Data standards make harmonization less challenging but may limit the amount of collected data. On the other hand, accepting a broader range of formats creates a harmonization hurdle. One solution is to accept formats that can be converted into multiple acceptable data structures (e.g., BIDS or NWB). While indiscriminate acceptance reduces time-consuming conversions by providers, it leaves the harmonization task to users.

Another topic of consideration is the potential introduction of batch effects when harmonizing and pooling data. While easily correctable, it is vital to consider when working with large and diverse datasets.

Requiring the BIDS and NWB standards allows the archives to have uniform data ready for secondary analysis. Structured data promote a consistent approach to data annotation, sharing, and storage and increase the reproducibility of results. Standardized naming conventions and file structures facilitate interoperability between software platforms and simplify analysis. Additionally, standards ensure that the data are available and accessible, even if the original researcher or institution no longer maintains the datasets.

A further obstacle to successful data sharing and secondary analysis is the limited guideline on data sharing protocols. Many researchers elect to keep their datasets or metadata private, while others upload incomplete sets, hindering the reanalysis efforts. Some repositories allow data owners to upload agreements that require co-authorship considerations, provide guidelines on crediting the sets, and outline stipulations before releasing the data. BRAIN Initiative's guidelines for data sharing have continued to improve by adapting to the needs of the data providers and users, removing ambiguity, and offering a policy that can be uniformly implemented to enhance data sharing in this field further.

In Canada, OBI mandates that all Integrated Discovery ("ID") Programs it funds provide data to Brain-CODE to foster collaboration and data sharing. Therefore, OBI's policy on data sharing states that the data produced through its funding should be accessible with minimal constraints in a responsible and timely manner[60].

Data redundancy, interoperability, performance optimization, maintenance and support, safeguarding against data loss, and comprehensive documentation enhance user navigation and help developers maintain and improve the system, which is fundamental to success. The importance of robust security mechanisms and a design conducive to growth complements these efforts.

**Future Directions**: This review aims to reflect the value and promise of neurophysiology platforms aggregating intracranial and imaging neurodata by highlighting their limitless benefits for the neuroinformatics community. Defining characteristics within and between DABI, DANDI, OpenNeuro, NEMAR, and Brain-CODE equip data users and owners with unique tools to better investigate, share, and analyze data.

One of the NIH BRAIN Initiative and OBI's objectives is to expand research through electrophysiology data sharing and secondary analysis, a model that has been exceptionally successful in other areas of research, including genomics and imaging neuroscience, where projects such as NIH-funded Alzheimer's Disease Neuroimaging Initiative (ADNI) have led to significant findings that resulted in new treatment protocols[61,62].

To further enhance the utility and impact of these platforms, several constraints should be considered when thinking about the future directions of data archives.

**Standardization Efforts**: While harmonization models may be crucial in the current landscape, they rely on post-collection processing, which can introduce potential distortions in the data. Establishing standardized data and behavioral task collection protocols between all archives should be the first step in this effort. Furthermore, limiting the number of possible file formats or proposing a universal format would eliminate the need for file harmonization. Implementing BIDS or NWB would solve the need for conventional file naming and organization. Lastly, developing tools that can aid conversion to a shared standard with limited input from researchers would reduce the time devoted to this effort.

Overall, the implementation and utilization of standardized protocols ensure uniformity in data collection, management, and analysis, reducing the need for extensive harmonization efforts. In addition, these protocols minimize variability and allow for data comparability across studies, fostering more robust and reliable integrative analyses. Standardization also aids in enhancing reproducibility and data sharing across platforms and databases.

**Analysis**: Developing reliable, cloud-based, collaborative, and accessible analysis platforms is essential to preserve researchers' time and computational resources. Such platforms would provide a valuable incentive for researchers to engage in secondary analyses, fostering scientific progress and collaboration.

Further, embracing the potential of artificial intelligence (AI) and machine learning (ML) technologies can significantly enhance data analysis and interpretation. Future directions should explore integrating AI and ML algorithms into neurophysiology platforms to automate data processing, identify patterns, and generate valuable insights.

Additionally, enhancing data visualization tools can aid researchers in more effective neurophysiological data analysis. The development of interactive and intuitive data visualization techniques can facilitate the interpretation of findings.

**Interoperability**: Investing in developing standardized data collection protocols and models that allow for interoperability of neurophysiology databases will facilitate large-scale integrated analytics without the need for extensive harmonization.

By promoting interoperability, neurophysiology platforms can facilitate cross-platform collaboration where researchers can share data, pool resources, and join projects, leading to more comprehensive studies with a broader impact.

Interoperability can enable the development of machine learning and artificial intelligence applications that simultaneously analyze data from multiple sources. AI-driven approaches can reveal complex patterns and associations that may not be apparent when analyzing data in isolation.

Furthermore, interoperability can be particularly valuable for longitudinal studies, where data must be collected and analyzed consistently. By promoting standardized data protocols across platforms, longitudinal studies would maintain data validity and reliability.

**Policy**: Continued efforts in developing and enforcing data-sharing policies are crucial to ensure compliance from both data owners and users. Stricter guidelines can uphold and reinforce ethical data sharing and reuse practices, cultivating a responsible and ethical scientific environment.

As data sharing becomes more prevalent, ensuring robust data privacy and security measures is crucial. Future directions should prioritize developing and implementing stringent data protection protocols to safeguard sensitive information and maintain the trust of data owners and participants.

**Long-Term Data Sustainability**: To preserve neurophysiology data for further research, long-term sustainability and accessibility of data archives are critical. Implementing strategies for data curation, preservation, and continuous maintenance will enable access to valuable data for future investigations.

Data archives improve how researchers across the scientific community can aggregate, analyze, and share multimodal data that contribute research to evidence-based medicine. Implementing these platforms permits large-scale experiments, ensures the reproducibility of findings, and enhances machine learning tools. Consequently, discovering new diagnostics and therapeutics that can be translated into patient care becomes a more realistic expectation and improves overall population health outcomes.

## Data Availability
Publicly available datasets stored on each of the repositories discussed can be downloaded from:
**DABI**: https://dabi.loni.usc.edu/search
**DANDI**: https://dandiarchive.org/dandiset
**OpenNeuro**: https://openneuro.org

**NEMAR:** https://nemar.org/dataexplorer
**Brain-CODE**: https://www.braincode.ca/content/open-data-releases

# Code Availability
**Notes about software discussed in the paper**
1. Note that DANDI utilizes the OMERO-Zarr, a software package for efficient storage and retrieval of large microscopy datasets.
2. For more information on Dandisets, see https://github.com/dandisets.
3. For more information on the DANDI API server, see https://api.dandiarchive.org
4. Note that DataLad datasets are standard git/git-annex repositories, and these tools may be used directly in cases where the DataLad tool is not desired or available.
5. While Jupyter alone is not optimal for use with electrophysiology data, it can be used with other Python libraries such as MNE-Python to load, preprocess, and plot example EEG data in a Jupyter notebook through vscode.
6. For NWB explorer, see http://nwbexplorer.opensourcebrain.org/ for more information.
7. For BioImagSuite/Viewer, see https://bioimagesuiteweb.github.io/webapp/viewer.html for more information.
8. For itk/vtk viewer, see https://kitware.github.io/itk-vtk-viewer/docs/ for more information.
**RAVE**
https://openwetware.org/wiki/RAVE:Install
**RStudio**
https://posit.co/download/rstudio-desktop/

**Acknowledgments.** We are grateful for contributions from Joyce Gong (University of Southern California, Los Angeles CA), Heena Cheema (Ontario Brain Institute, Toronto ON, Canada), Derek Eng (Ontario Brain Institute, Toronto ON Canada), Fatema Khimji (Ontario Brain Institute, Toronto ON, Canada), Francis Jeanson (Datadex, Toronto ON, Canada), Bianca Lasalandra (Indoc Research, Toronto ON, Canada), Emily Martens (Indoc Research, Toronto ON, Canada), Anthony L Vaccarino (Indoc Research, Toronto ON, Canada), Mojib Javadi (Indoc Research, Toronto ON, Canada), Kirk Nylen (Ontario Brain Institute, Toronto ON, Canada)

**Funding.** This study was supported by the National Institutes of Health (NIH) under Award Number R24MH114796 (DABI), R24MH117295-04 (DANDI), R24MH117179-03S2 (OpenNeuro), R24MH120037-01 (NEMAR), R24MH117529-01 (RAVE), R01MH111417 (BIDS), R24MH116922 (NWB).
This study was also supported by The National Institute of Neurological Disorders and Stroke (NINDS) under award number R37NS21135 (BIDS).

## Author Contributions

## Competing Interests

**DABI Affiliated Researchers:** Priyanka Subash, Alex Gray, Misque Boswell, Samantha L. Cohen, Rachael Garner, Sana Salehi, Calvary Fisher, Samuel Hobel, Nader Pouratian, Arthur W. Toga, and Dominique Duncan; **DANDI and NWB Affiliated Researchers:** Satrajit Ghosh, Yaroslav Halchenko, Benjamin Dichter; **OpenNeuro Affiliated Researchers:** Russell A. Poldrack, Chris Markiewicz; **BIDS Affiliated Researchers:** Dora Hermes; **NEMAR Affiliated Researchers:** Arnaud Delorme, Scott Makeig; **Brain-CODE Affiliated Researchers:** Brendan Behan, Alana Sparks; **RAVE Affiliated Researchers:** Zhengjia Wang, John Magnotti, Michael Beauchamp.